\documentclass[prx,aps,superscriptaddress,twocolumn,amsmath,amssymb,floatfix]{revtex4-2} 

\usepackage[breaklinks,colorlinks = true,linkcolor = magenta,urlcolor=magenta,citecolor=red,hypertexnames=false]{hyperref}
\usepackage{graphicx}
\usepackage{overpic} 
\usepackage{multirow}

\usepackage{dcolumn}
\usepackage{bm}
\usepackage{color}
\usepackage{tikz}

\usepackage{tabularx}
\usepackage[dvipsnames]{xcolor}
\usepackage{lipsum}
\usepackage{braket}

\usepackage{silence}
\allowdisplaybreaks
\def\be{\begin{equation}}
\def\ee{\end{equation}}

\makeatletter
\newsavebox{\@brx}
\newcommand{\llangle}[1][]{\savebox{\@brx}{\(\m@th{#1\langle}\)}%
  \mathopen{\copy\@brx\mkern2mu\kern-0.9\wd\@brx\usebox{\@brx}}}
\newcommand{\rrangle}[1][]{\savebox{\@brx}{\(\m@th{#1\rangle}\)}%
  \mathclose{\copy\@brx\mkern2mu\kern-0.9\wd\@brx\usebox{\@brx}}}
\makeatother

\begin{document}
      \title{Continuously varying exponents in the distribution of waiting times in the\\ symmetric exclusion process on a percolation cluster}

\author{Arpan Chatterjee}
\affiliation{Tata Institute of Fundamental Research, Hyderabad 500046, India}
\author{Kabir Ramola}
\affiliation{Tata Institute of Fundamental Research, Hyderabad 500046, India}
\author{Deepak Dhar}
\affiliation{International Centre for Theoretical Sciences,\\ Tata Institute of Fundamental Research, Bengaluru 560089, India}

\date{\today}

\begin{abstract}
We study the waiting-time distribution of hard-core interacting particles in the symmetric exclusion process on one- and two-dimensional lattices with side branches attached to each lattice site in the steady state. We use numerical simulations together with an approximate analytical treatment of particles trapped in side branches in the steady state. Such a two-dimensional carpet serves as a simplified model for trapping of a supercritical percolation cluster. At high particle densities, the system exhibits strong dynamical heterogeneity, with the distribution of logarithms of waiting times developing well-separated peaks corresponding to particles trapped at different depths from the backbone. We show that the probability that a tagged particle occupies the same position at time $t_0+t$ as at time $t_0$ decays algebraically as $t^{-\omega}$, where the exponent $\omega$ varies continuously with particle density for densities above a threshold value $\rho^*<1$. We also investigate the correlations between successive waiting times along the trajectory of a tagged particle.
\end{abstract}
\maketitle

\section{Introduction}
\label{sec:introduction}

The problem of fluid flow in porous media is a classic problem in statistical physics. Transport and diffusion in disordered media have been studied extensively for many decades because of their relevance to systems ranging from petroleum reservoirs to intracellular transport~\cite{P.Levitz_1997,PhysRevE.75.045103,PhysRevLett.94.248701,RevModPhys.45.574,PhysRevLett.27.1722}. Initially introduced as a model for studying diffusion of gases in coal~\cite{diffusion_in_coal}, the percolation model serves as a paradigm for studying transport in disordered media~\cite{PG_gennes,Havlin01011987,BOUCHAUD1990127,Drift_trapping_in_biased_disordered_lattice,M_Barma_1983,10.1214/18-AIHP901,PhysRevLett.134.027102,Broadbent_Hammersley_1957,PhysRevE.53.4187}. Diffusion and transport are commonly studied by considering random walks within the infinite percolation cluster paradigm~\cite{PhysRevB.29.511,PG_gennes,PhysRevE.65.021112}. We distinguish between the sites of a spanning cluster on the backbone, and on the dead ends or side branches.
The symmetric exclusion process (SEP)~\cite{Kumar_2020,MALLICK201517,Mishra_2023,PhysRevE.100.032136,Derrida2009,f841a5ed-f648-3227-bf50-ba46dbcbbc81,PhysRevLett.118.160601, Mason2023} on a supercritical percolation cluster provides a minimal model for diffusion in disordered and crowded environments. The random cluster connectivity captures the complex geometry of porous or biological networks~\cite{RevModPhys.74.47,Jeong2000}, while hard-core interactions~\cite{PhysRevLett.51.1729} account for the excluded-volume constraint that particles cannot overlap. Such a model is relevant to ionic or molecular transport through nanoporous materials, zeolites, and crowded intracellular environments, where particles undergo unbiased diffusion within the random geometry of the accessible connected region of space.

\begin{figure}
    \centering
    \includegraphics[width=1\linewidth]{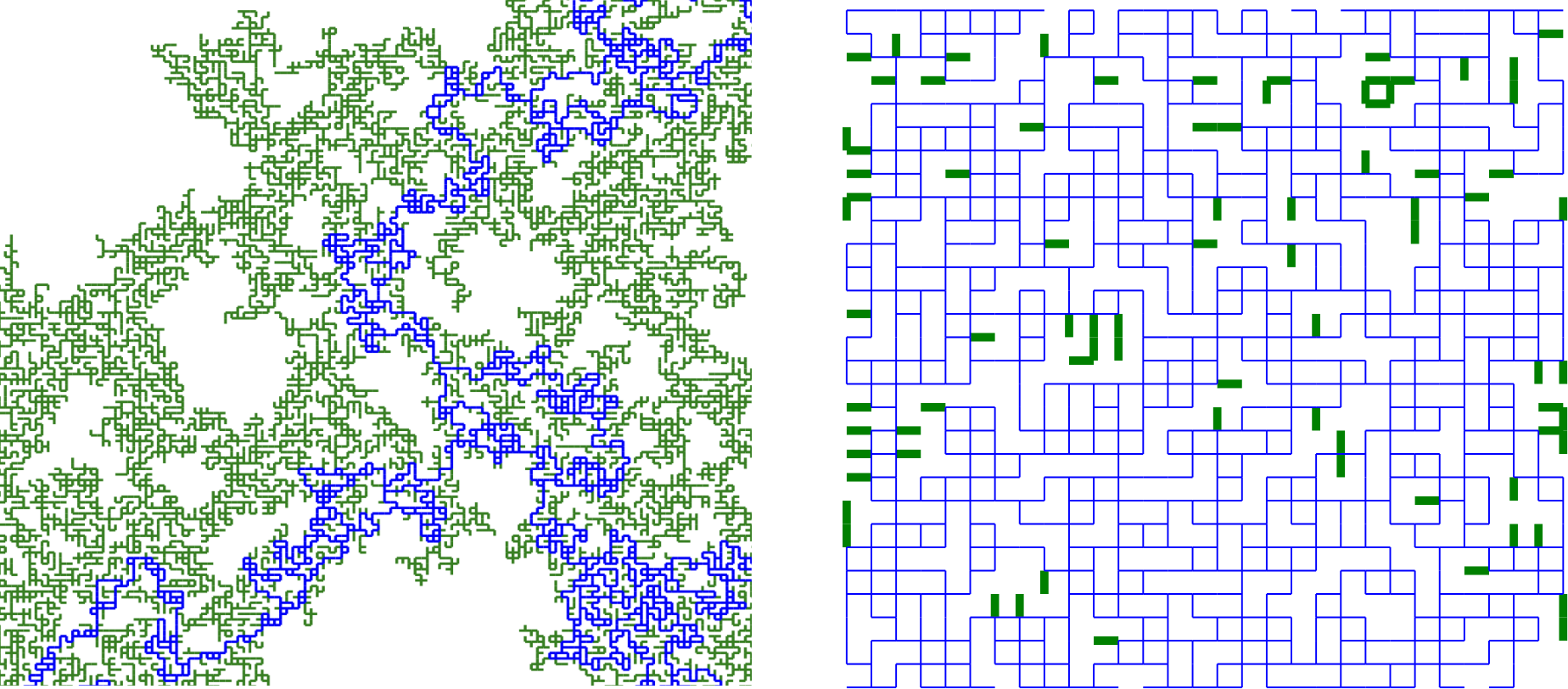}
    \caption{Representative spanning clusters. The left panel shows a critical bond percolation cluster generated on a $200\times 200$ lattice, with the backbone highlighted in blue and side branches highlighted in green. The right panel shows a system-spanning cluster on a $30\times 30$ lattice at $p=0.70$.}
    \label{fig:snap_perc_cluster}
\end{figure}

Recently, Iyer \textit{et al.}~\cite{PhysRevLett.134.027102} studied the asymmetric exclusion process (ASEP)~\cite{Dhar1987,PhysRevLett.68.725,PhysRevE.98.052122} on a percolation cluster. The waiting time, $t_w$, is defined as the time a particle spends on the same branch since leaving the backbone. Their principal finding was that, in the presence of sufficiently strong external bias, the distribution of $\log t_w$ develops multiple well-separated peaks. It is natural to ask if this behavior persists even in the absence of an external field? In this paper, we show that the answer is yes. Specifically, for a particle at steady state, what is the probability that its waiting time on the current branch, measured from the most recent visit to the backbone, exceeds $t_w$? Even in the absence of an external bias, sufficiently high densities generate a ``topological bias''~\cite{PhysRevE.82.066109,PhysRevB.34.8129} that pushes particles into side branches, again leading to well-separated peaks in the distribution of $\log t_w$.
Although the motion of a tagged particle on a supercritical ($p>p_c$) percolation cluster is diffusive in steady state, the distribution of residence times within a branch exhibits nontrivial behavior and develops a power-law tail at high densities. We begin by studying a more tractable geometry, which we refer to as a \emph{carpet}: a two-dimensional lattice where each site is connected to a linear side branch of length $H$. First, understanding the behavior of the waiting-time distribution in this simpler setting, we then discuss how the distribution is modified in the case of SEP on a supercritical percolation cluster~\cite{FERRARI199189,Rammal1983RandomWO, alexander:jpa-00232103, 10.1214/EJP.v10-240}.

On a regular $d$-dimensional lattice, a tagged particle has diffusive behaviour $\langle r^2\rangle \sim t$, while the return probability scales as $P_{\mathrm{ret.}}(t)\sim t^{-d/2}$. In this paper, we show that on a supercritical percolation cluster in $d$ dimensions, at high densities the tagged particle remains diffusive with $\langle r^2\rangle \sim t$, but the return probability obeys
$P_{\mathrm{ret.}}(t)\sim t^{-{\omega}}$, 
where $\omega$ continuously varies with particle density $\rho$ and correlation length of the cluster $\xi$.

There has been considerable work on interacting particle systems in disordered geometries~\cite{Ramaswamy_1987,PhysRevLett.134.027102,Derrida1998,Spohn1991,Hilhorst_1998,BRUMMELHUIS1989575,PhysRevE.49.4946}. For large $t$, the return-to-origin probability in the symmetric exclusion process on a percolation cluster is dominated by contributions from particles that remain confined to the same side branch throughout the entire interval. Furthermore, we show that this exponent is a continuous function of density, $\rho$ and bond occupation probability, $p$. On a supercritical percolation cluster (with $P(h) \sim e^{-h/\xi}$), the probability that a particle is still found at the site it occupied at time $t_0$ after an elapsed time $t$ decays as $t^{-\omega}$, where $\omega=-\frac{1}{\xi \ln(1-\rho)}$.
\begin{figure}[t]
    \centering
    \begin{minipage}{0.9\linewidth}
        \centering
        \includegraphics[width=\linewidth]{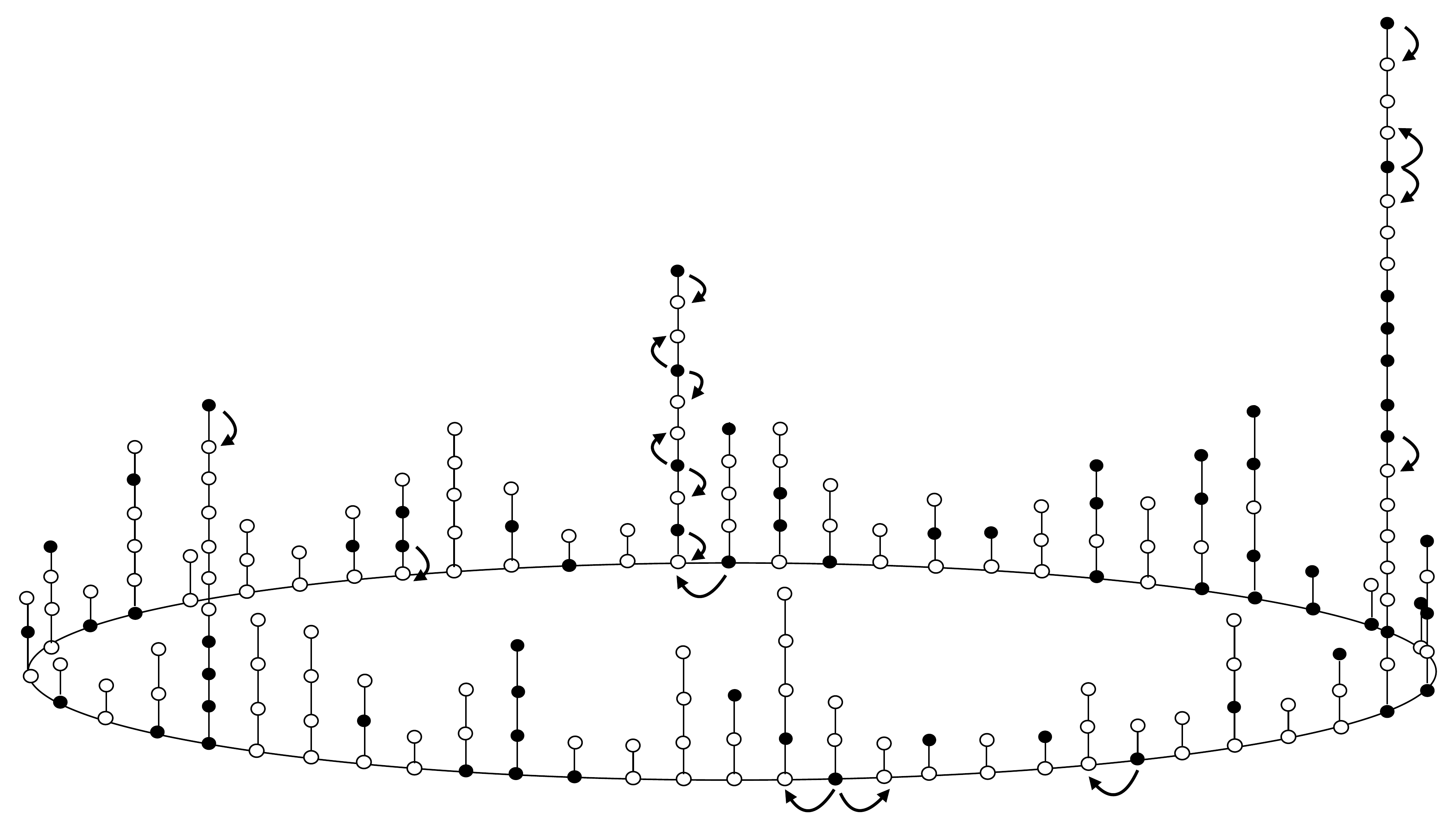}
    \end{minipage}
    \hfill
    \begin{minipage}{0.9\linewidth}
        \centering
        \includegraphics[width=\linewidth]{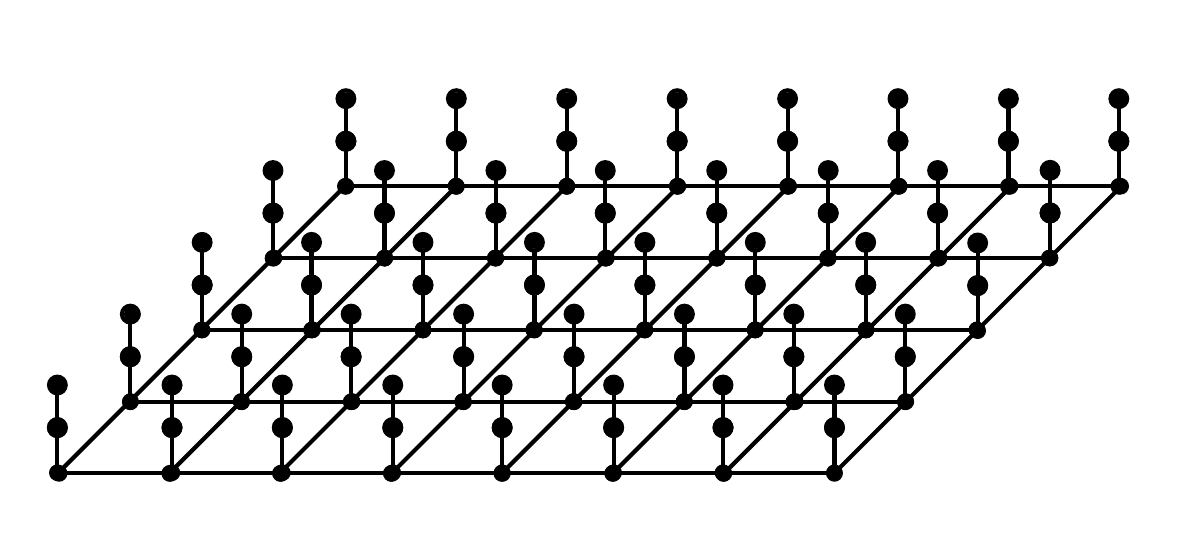}
    \end{minipage}
    \caption{Schematic diagrams of the models studied. \textbf{Top}: irregular comb with exponentially distributed side-branch lengths. The particles (dark circles) perform symmetric exclusion moves (arrows). For clarity, only a subset of the possible particle hopping moves is shown.
\textbf{Bottom}: The carpet graph with side branches of depth $H=2$.}
    \label{fig:carpet_exp_comb}
\end{figure}

\begin{figure}
    \centering
    \includegraphics[width=1\linewidth]{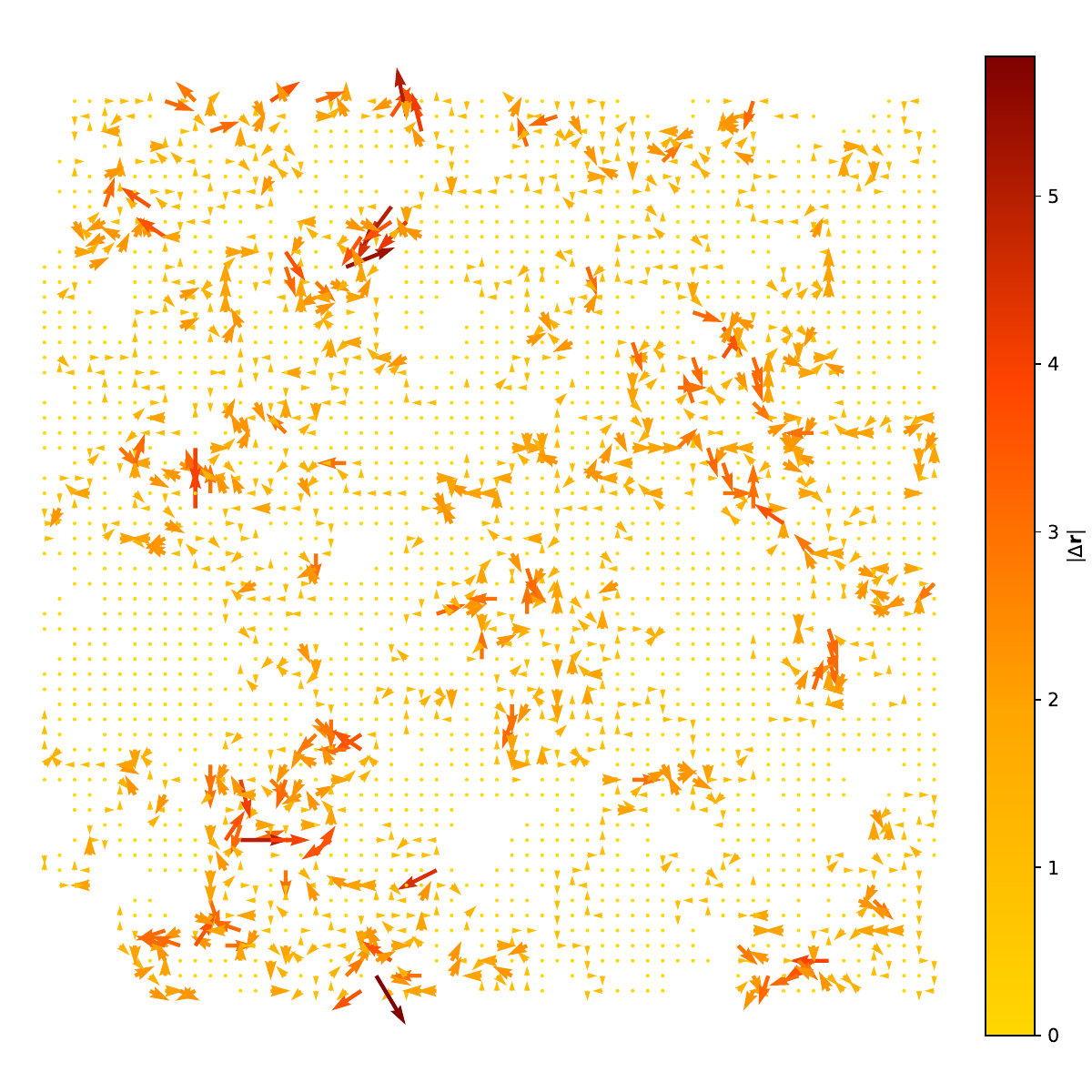}
    \caption{Displacement field on a percolation cluster (at $p=0.56$) revealing dynamical heterogeneity. Arrows represent particle displacements $|\Delta r|$ over a fixed time interval of $\Delta t =100$ on a $60 \times 60$ lattice at packing fraction $\rho=0.96$, colored by displacement magnitude.~The color bar indicates the magnitude of the displacements, ranging from $0$ (yellow) to $>5$ (red).~Yellow highlights the immobile regions in the percolation cluster, while red regions highlight maximally mobile particles.~The vast majority of particles remain frozen ($|\Delta r|\approx0$), forming a heterogeneous immobile background, while a small fraction execute large cooperative displacements organized into string-like mobile clusters. 
    }
    \label{fig:disp_field}
\end{figure}

%%%%%%%%%%%%%%%%%%%%%%%%%%%%%%%%%%%%%%%%%%%%%%%%%%%%
\section{The Model}
\label{sec:models}
To understand the effect of high density on the distribution of waiting times, it is first useful to study the problem on a lattice without disorder. We consider a square lattice with a linear side branch of $H$ sites attached to every lattice site. Every vertex can hold at most one particle; we refer to this geometry as the ``carpet''. The position of a vertex is given by three integers $(x,y,z)$. We will call the vertices at $z=0$ the floor of the carpet [Fig.~\ref{fig:carpet_exp_comb}].
Periodic boundary conditions along the $x$ and $y$ axes are imposed only on the floor of the carpet. Vertices with $z>0$ form the side branches. Particles can move from one branch to another only by moving along the floor.

\subsection{Symmetric Exclusion Process}
We first discuss the Symmetric Exclusion Process (SEP) on the carpet. At $t=0$, each vertex is initialized with a particle with probability $\rho$, and particles move obeying the symmetric exclusion process. In simulations, we keep a list of active bonds; by active bonds, we mean those bonds where the occupancy at the two ends is different. We pick an active bond at random and flip the particle-hole pair to a hole-particle pair with rate $1$ (see Fig.~\ref{fig:carpet_exp_comb}). The list of active bonds is updated after every move. The algorithm used here is rejection-free. The algorithm avoids rejected moves and does not slow down even when the density of particles is very close to $1$, allowing us to simulate densities near unity. In this text, we will call this algorithm the ``Monte Carlo'' update from now on. The model is defined in continuous time; in our implementation, a single move of a particle increments the time by $1/A$, where $A$ is the number of active bonds at that instant. For each particle, we keep track of the last time step when the particle was on the floor. As particles move throughout the carpet, we measure the time a particle spends stuck in a side branch before returning to the floor, which we call the ``residence time''. The waiting time, in contrast, is the elapsed time since a particle entered a side branch and has not yet returned to the floor.

\subsection{Lifted SEP and Toom Model}

%%%%%%%%%%%%%%%%%%%%%%%%%%%%%%%%%%%%%%%%%%%%%%
\begin{figure}[t]
    \centering
    \includegraphics[width=0.8\linewidth]{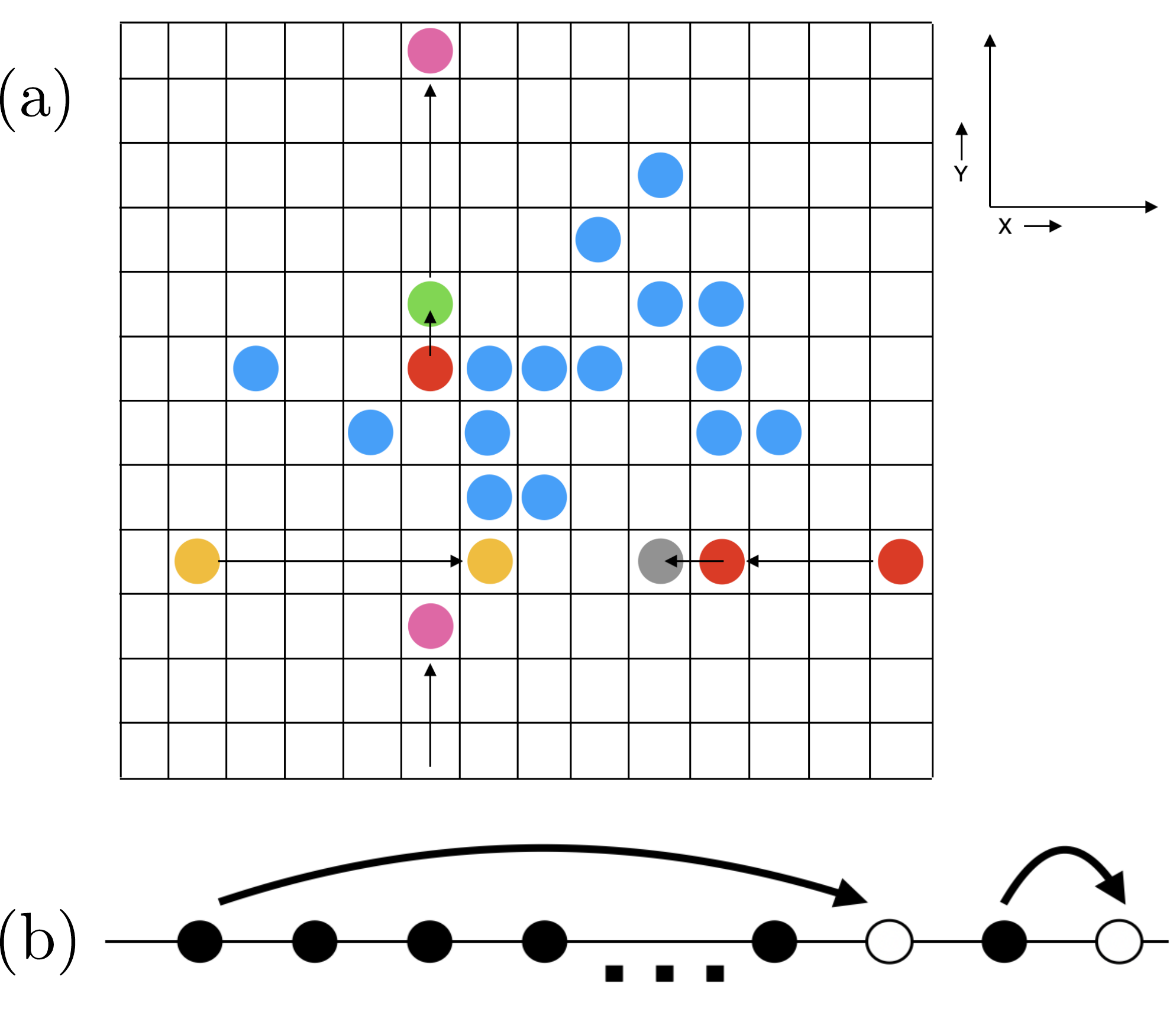}
    \caption{(a) Schematic illustration of several lifted-SEP (LSEP) update events on the floor of the 2D carpet. Consider first the red--green pair: at time $t$, the red particle is selected for an LSEP move in direction $+\hat{y}$ with displacement $D = 8$. The red particle immediately collides with the green particle, which then propagates freely for three steps before colliding with the pink particle. The pink particle continues for the remaining three steps, wrapping once around the periodic lattice to complete the move. A second example involves the yellow particle, chosen with $D = 5$ in the $+\hat{x}$ direction, which moves freely for the entire displacement. Additional example moves are also shown. Three directions are depicted here for clarity, though in general all four lattice directions are possible.
    (b) A possible Toom's move on the floor of a one-dimensional lattice, illustrating a long-range jump.}
    \label{fig:lsep_schm}
\end{figure}
%%%%%%%%%%%%%%%%%%%%%%%%%%%%%%%%%%%%%%%%%%%%%%%
In order to better understand the nature of the correlations, we also study two variants of the SEP; the Lifted SEP~\cite{LSEP,PhysRevX.14.041035} and the Toom model. The Lifted SEP has the same equilibrium marginal distribution as the SEP, but exhibits faster relaxation. This move is defined as follows: pick a particle on the floor at random and choose a step size $D\gg1$. Pick a random direction among $+\hat{x}, -\hat{x}, +\hat{y}, -\hat{y}$. Move the chosen particle in that direction as many steps as possible without colliding with another particle. When a collision occurs, the other particle becomes the moving particle and this loop continues until the total displacement equals $D$, at which point the move ends (see Fig.~\ref{fig:lsep_schm}(a)).

We consider another variation of the model studied by Toom~\cite{MR469584} and Derrida~\textit{et al.}~\cite{Derrida_1991}. They originally introduced this model as an effective one-dimensional description of interface dynamics in the two-dimensional Toom cellular automaton~\cite{Derrida_1991,PhysRevLett.67.165}. At each update, a particle on the floor is chosen uniformly at random, and the nearest hole to its right is identified. If a particle-hole pair is found at positions $m$ and $n$ ($m < n$), their occupations are exchanged (Fig.~\ref{fig:lsep_schm}(b)). This rule is adapted from the Toom dynamics introduced by Derrida~\textit{et al.}~\cite{Derrida_1991}, where a randomly selected spin is exchanged with the nearest spin of opposite sign to its right. In the original model, if no such spin exists, the selected spin is flipped. Our implementation omits this self-flipping step to conserve the particle density. 

\subsection{SEP on Percolation Clusters}
We next show that the power-law envelope of the waiting-time distribution persists on supercritical percolation clusters. To this end, we study particles undergoing SEP dynamics on actual percolation clusters. We consider a two-dimensional square lattice and construct the complete list of nearest-neighbor bonds. Each bond is retained with probability $p$, or equivalently, a fraction $(1-p)$ of the bonds is randomly removed. The remaining bonds define a realization of a bond percolation cluster. Backbone and side branches are identified using a depth-first search. We start with the full spanning cluster. We then iteratively remove from the backbone all bonds that have only one neighbor bond. This procedure is repeated using the updated backbone until no such bonds remain. More complex side branches can then be identified as connected structures attached to the backbone through single bond~\cite{doi:10.1142/S0129183103005509,HJHerrmann_1984}. Once this decomposition has been obtained, the mass of each side branch is computed, and used to determine the distribution of mass and length of side branches. Their statistical properties have been studied extensively in the literature \cite{PORTO199996,PhysRevE.51.2632,PhysRevE.89.012120,StaufferAharony1992}. At the percolation threshold in two dimensions, the branch mass distribution follows a power law $P(M) \sim M^{-1.83}$, while the branch length distribution scales as $P(L) \sim L^{-2.42}$~\cite{PORTO199996}.
In parallel, transport on critical percolation clusters has also been investigated extensively. Owing to the fractal geometry of the spanning cluster, particles exhibit anomalous diffusion, with the mean square displacement growing sub-linearly in time~\cite{Havlin01011987,M_Barma_1983,Rammal1983RandomWO,10.1214/EJP.v10-240}. Away from $p_c$, the distribution of side branches is an exponential distribution.

Transport of particles occurs through the backbone of the spanning cluster, whereas side branches act as traps for particles (see Fig.~\ref{fig:snap_perc_cluster}). In a similar fashion, we define the ``waiting time'' in a side branch within a percolation cluster to be the time a particle spends stuck in the side branch since its last step leaving the backbone.

%%%%%%%%%%%%%%%%%%%%%%%%%%%%%%%%%%%%%%%%%%%%%%%%%%%%
\section{Preliminaries and Single Branch Analysis}
\label{sec:preliminaries}
In this section, we first consider a single branch, coupled to a memoryless particle bath. We begin by describing the microscopic dynamics and the corresponding steady state, followed by a description of the algorithm used.
\subsection{Preliminaries}
\subsubsection{Model and Steady State}
We consider an infinite percolation cluster at $p>p_c$. Within the cluster, we initialize the particles randomly with particle density $\rho$, all performing moves obeying the symmetric exclusion process. On a percolation cluster near the critical point, the side branches have loops and a fractal structure, but away from the critical point most of the side branches are $1\text{D}$ chains (see Fig.~\ref{fig:snap_perc_cluster}). Hence, we first consider a single linear side branch in the cluster where particles enter and exit the branch through one end.

%%%%%%%%%%%%%%%%%%%%%%%%%%%%%%%%%%%%%%
\begin{figure}[t!]
    \centering
    \includegraphics[width=0.9\linewidth]{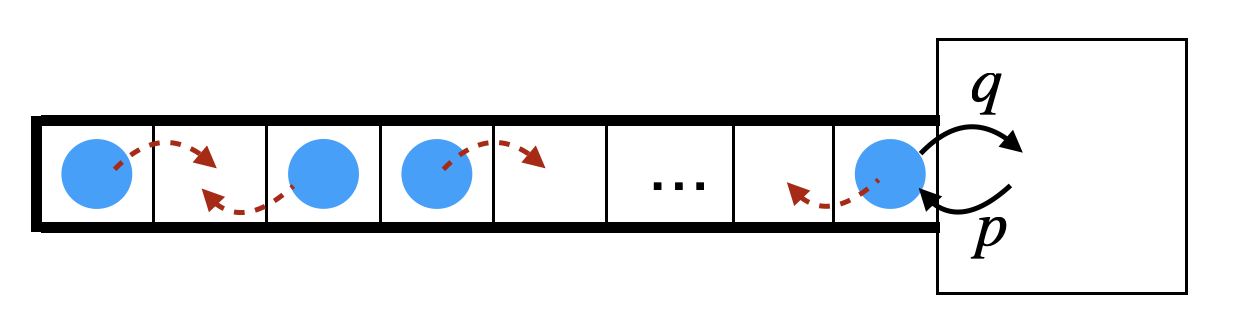}
    \caption{Schematic of a single branch illustrating the microscopic dynamical rules. Particles enter and leave the branch through the rightmost site, which is connected to a particle reservoir maintained at density $\rho$. Entry and exit occur with rates $p=\rho$ and $q=1-\rho$, respectively. The transitions indicated by dotted arrows occur with unit rate.}
    \label{fig:single_branch}
\end{figure}
%%%%%%%%%%%%%%%%%%%%%%%%%%%%%%%%%%%%%%

%%%%%%%%%%%%%%%%%%%%%%%%%%%%%%%%%%%%%%
\begin{figure*}
    \centering
    \includegraphics[width=1\linewidth]{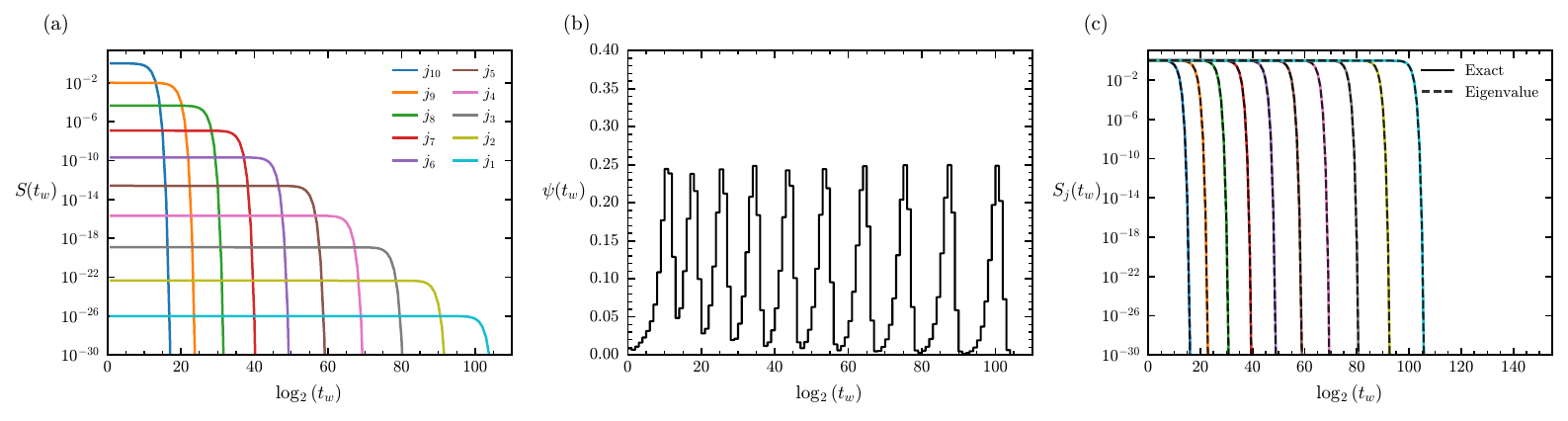}
    \caption{(a) Cumulative waiting time distribution of particles inside a branch of depth $H=10$ at $\rho=0.999$.~The variables $j_n$ label the conditional probability that at least $n$ particles survive in the branch up to time $t_w$. (b) Probability density $\psi(t_w)$ of waiting time $t_w$ for a particle in a side branch of depth $H=10$ at density $\rho=0.999$. (c) Comparison of $S_j(t)$ calculated using the restricted matrix method, $W_j$ (Exact), and the smallest eigenvalue approximation $e^{-\lambda_j t_w}$.}
    \label{fig:waiting_time_summary}
\end{figure*}

%%%%%%%%%%%%%%%%%%%%%%%%%%%%%%%%%%%%%%%%%%%%%%%%%%%%

\subsection{Motion in a Single Branch}
\label{sec:single-branch}

For a single branch of length $H$ connected to a particle bath, at every instant a bond is chosen with uniform probability, and the particle-hole pair is flipped to a hole-particle pair and vice versa with rate $1$. If a bond that lies at the mouth of the branch is picked, a hole (particle) is converted to a particle (hole) with rate $p~(q)$ where $p=\rho$ and $q=(1-\rho)$ (see Fig.~\ref{fig:single_branch}). The steady state of the symmetric exclusion process has a product measure; each tagged particle spends an equal fraction of time at each site.

As particles move along the branch, the state of the branch transitions from one state to another among the $2^H$ valid states. For the Markov process describing the branch states, the evolution is given by the master equation,
\begin{equation}
    \label{eq:master_eq}
    \frac{d}{dt}\ket{P(t)} = \mathbf{W}\ket{P(t)} \, 
\end{equation}
where $\ket{P(t)}$ is the probability density of branch states at time $t$, represented by a vector of dimension $2^H$. In the steady state, we ask for the probability that a particle is still in the branch at time $t$ after entering at time $t=0$.

\begin{figure*}
    \centering
    \includegraphics[width=1\linewidth]{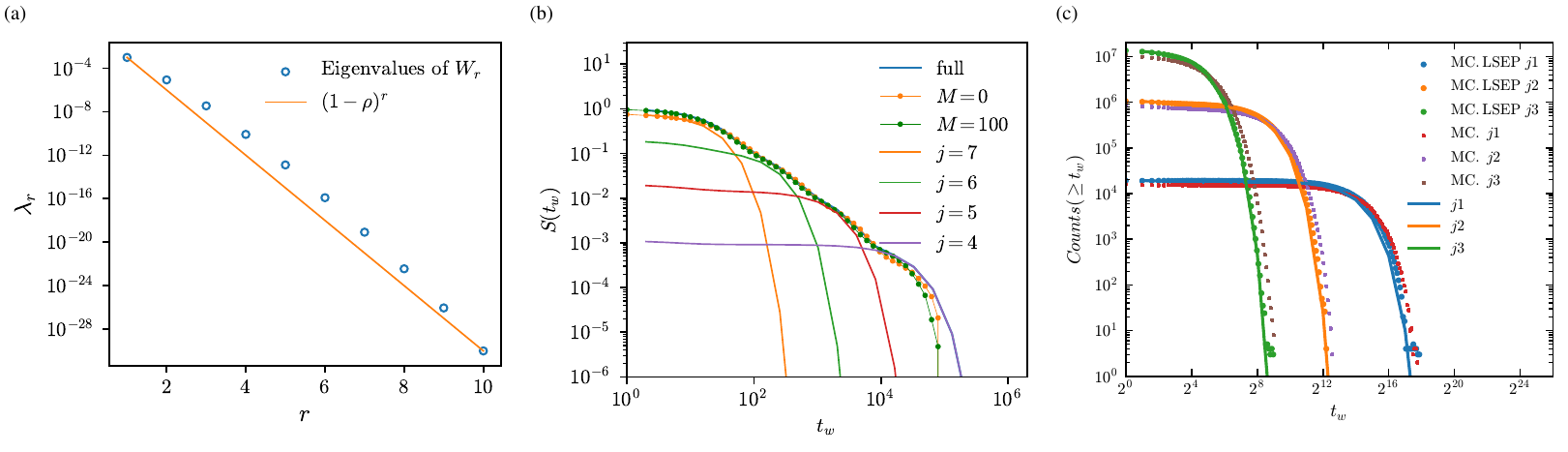}
    \caption{(a) Variation of the smallest non-zero eigenvalues of the restricted matrices compared with minimum number of particles $r$ in the branch. (b) Exact calculation of survival probability of particles in a side branch of depth $H=7$ at $\rho=0.96$ compared against simulations with Lifted--SEP rules ($M=100$) and without LSEP rules ($M=0$) on a two-dimensional carpet. $M$ signifies that for every Monte Carlo move, $M$ LSEP moves are performed on the floor of the carpet. The variables $j=1, 2, 3,...$ labels the conditional probability that at least $1, 2, 3, ...$ particle(s) survive in the branch up to time $t$. The label `full' represents the sum of the individual curves corresponding to $j=4 \text{ to } 7$.
    (c) Histogram of waiting time distribution measured on the carpet with LSEP rules (MC.LSEP) and without LSEP rules (MC.) on a two-dimensional carpet.~Introducing LSEP steps weakens the correlations between successive waiting times, improving the agreement between the single branch approximation and simulation.~The data are for branch depth of $H=3$, particle density $\rho=0.96$ and $M=1000$.}
    \label{fig:eig_surv_counts}
\end{figure*}
%%%%%%%%%%%%%%%%%%%%%%%%%%%%%%%%%%%%%%%%%%%%%%%%%%%%

One defines a subspace $\mathcal{H}_r$ containing all configurations with at least $r$ particles. The system evolves from one configuration to another as time progresses. The probability that the last added particle is still in the system is equal to the probability that a particle entering the side branch when there are $(r-1)$ particles already present in the side branch. Given the transition matrix $\mathbf{W}$, we construct a restricted matrix $\mathbf{W}_r$ acting on the subspace $\mathcal{H}_r$ by deleting rows and columns corresponding to configurations containing fewer than $r$ particles.

The survival probability $S(t|r)$ is defined as the probability that there are at least $r$ particles in the side branch that remain in the side branch up to time $t$,
\begin{equation}
    S(t|r) = \bra{1, 1, .., 1}e^{\mathbf{W}_r t}\ket{g(r)}
    \label{eq:surv_prob}
\end{equation}
where $\ket{g(r)}$ is the projection of the steady-state vector onto this subspace. The entries of this vector are non-zero for those configurations that have precisely $r$ particles with one particle at the rightmost site. The total cumulative waiting-time distribution is the weighted sum of the individual survival probabilities, given by
\begin{equation}
    S(t) = \sum_{r} S(t \mid r)\, w(r)
    \label{eq:surv}
\end{equation}
where $w(r)$ is the probability that there are $r$ particles in the branch at $t=0$.
We define one update step as $\ket{P(t+1/B)} = (\mathbf{1} + \frac{\mathbf{W}}{B})\ket{P(t)}$, and increment the clock time by $1/B$, where $B$ is the number of bonds. Then the probability vector after clock time $t$ is
\begin{equation}
    \ket{P(t)} = \left(\mathbf{1} + \frac{\mathbf{W}}{B}\right)^{Bt}\ket{P(0)}.
\end{equation}

In Fig.~\ref{fig:waiting_time_summary}(a), we have calculated the survival probability of a particle in a side branch of depth $H=10$ at density $\rho=0.999$. Each curve is labeled by $j$, denoting the minimum number of particles present in the branch. For instance, $j_1$ corresponds to at least one particle in the branch, $j_2$ to at least two, and so on. The calculation uses discrete-time evolution at times $t=2^n$, obtained by repeated matrix squaring: the evolution matrix at step $n$ is simply the square of that at step $n-1$, since for any matrix $M$, $M^{2^n} = \left[M^{2^{n-1}}\right]^2$. Further, we have calculated the probability density of waiting times in Fig.~\ref{fig:waiting_time_summary}(b). In the regular comb, it is equally likely for a particle to be at any depth in the steady state, which makes the amplitudes of the peaks equal in Fig.~\ref{fig:waiting_time_summary}(b).

% {\color{blue} 
Calculating the survival probability exactly requires arbitrary-precision arithmetic to avoid numerical underflow or overflow, which is computationally prohibitive. Instead, we investigate whether the long-time behavior of the survival probability can be accurately captured by a subset of the eigenvalues of the restricted matrix. In general, the survival probability $S(t\vert{}j)$ for having at least $j$ particles can be expressed as a sum of exponentials over all eigenvalues $\lambda_\alpha$ of the corresponding restricted matrix $\mathbf{W}_j$:
\begin{equation}
S(t\vert{}j) = \sum_{\alpha}C_{\alpha}e^{-\lambda_{\alpha}t}.
\end{equation}
Evaluating this full eigenvalue expansion requires knowledge of the projection coefficients $C_{\alpha}$, subject to the normalization constraint $\sum_{\alpha}C_{\alpha} = 1$. However, we observe that at sufficiently high densities, the smallest non-zero eigenvalue alone is sufficient to reproduce the asymptotic, long-time behavior of the survival probability. Hence, the survival probability curve $S_j(t)$ can be approximated as:
\begin{equation}
S_j(t) \approx e^{-\lambda_j t}.
\end{equation}
In Fig.~\ref{fig:waiting_time_summary}(c), we compute the survival probability using the full $\mathbf{W}$ matrix evolution and compare it with the single-eigenmode approximation of the restricted matrix. The agreement between the exact matrix calculation and the single-eigenmode approximation systematically improves at higher particle densities.

Because the restricted matrices are of size $2^H \times 2^H$, where $H$ is the depth of the side branch, numerical diagonalization remains computationally expensive for deep branches. This raises a natural question: can the smallest non-zero eigenvalue be approximated by a simple scaling form in terms of the particle density? 
It is easy to see that the smallest non-zero eigenvalue scales with the density as $\lambda_r \simeq (1-\rho)^{r}$.
In Fig.~\ref{fig:eig_surv_counts}(a), we plot the smallest non-zero eigenvalue $\lambda_r$ of the restricted matrix $\mathbf{W}_r$ against the scaling of density $\rho$ to verify this relationship. Furthermore, using this scaling relation, we can directly estimate the characteristic survival timescale $t_r$ for $r$ particles in a side branch. Defining $t_r$ as the time required for the survival probability $S(t)$ to decay to a given threshold $C \in (0, 1)$, we have:
\begin{align}
    S(t_r) = C \approx e^{-\lambda_r t_r};~~~
    t_r = \frac{\ln(1/C)}{\lambda_r}.
\end{align}
Substituting the scaling form $\lambda_r \sim (1-\rho)^{r}$ yields the exponential hierarchy of trapping times:
\begin{equation}
    t_r \sim \left(\frac{1}{1-\rho} \right)^r.
\end{equation}
Thus, each additional particle in the side branch multiplies the characteristic escape time by a factor of $\frac{1}{1-\rho}$. Therefore, $\ln t_r \sim r \ln(\frac{1}{1-\rho})$, directly explaining the occurrence of equally spaced peaks in the probability density profile $\psi(t)$ when plotted on a logarithmic time scale (see Fig.~\ref{fig:waiting_time_summary}(b)).
% }

\section{Extension to Carpets and Percolation Clusters}
\label{sec:extension}
We can extend the treatment of a single branch to any finite cluster of sites, replacing the effect of the rest of the lattice by an independent particle bath at each dangling side branch.

\subsection{Extension to Carpets}

We have performed simulations on a two-dimensional regular carpet and compared the results with the single-branch approximation (see Fig.~\ref{fig:eig_surv_counts}(b)-(c)) and studied the deviations from the single branch approximation. We observe a very similar qualitative match between the single-branch approximation and simulation results. The staircase distribution exhibits step spacing along the $\log(t)$ axis on the order of $\log(\frac{1}{1-\rho})$. This deviation is due to the correlations present among the sojourn times. At densities, $\rho >0.95$, it is difficult to eliminate all such correlations; however, in the case of \emph{lifted} SEP dynamics on the floor of the carpet, the correlations are weaker. In our simulations, we use two types of updates for moving the particles. Type 1 is the ``Monte Carlo'' update, and Type 2 is the ``Lifted SEP''. For every update of Type 1, there are $M$ Type 2 updates, where $M\gg1$. The qualitative behavior of the waiting-time distributions is not affected, but the agreement between the single-branch approximation and simulation data improves (see Fig.~\ref{fig:eig_surv_counts}(b)-(c)). Clearly, in the limit $M \to \infty$, the correlation between sojourn times can be eliminated completely.

\subsection{Supercritical percolation cluster}

Finally, we turn our attention to the waiting time distribution on a supercritical percolation cluster. On a supercritical percolation cluster with a correlation length $\xi$, the geometric distribution of the side branches fundamentally alters the asymptotic transport properties of the system. For $p > p_c$, the probability of finding a side branch of depth $h$ decays exponentially due to the quenched disorder of the cluster, scaling as $P(h) \sim \exp(-h/\xi)$. As established in our single-branch analysis, the maximum relaxation time (or exit time) for a particle in a branch of depth $h$ is governed by the slowest eigenmode, yielding a characteristic trapping time $t \sim (1-\rho)^{-h}$. 

By inverting this relation,
\begin{equation}
h \simeq 
      \frac{\ln t}{\ln\!\left(\frac{1}{1-\rho}\right)},
\end{equation}
we map the spatial distribution of branch depths onto the temporal distribution of waiting times. It is convenient to work with  $\psi(t)$, the waiting-time density per logarithmic interval.

Therefore for continuous branch depths, the envelope of waiting-time probability density is
\begin{equation}
\psi(t)\simeq
\exp\!\left[
\frac{\ln t}
{\xi\ln(1-\rho)}
\right]
\sim
t^{-\omega}.
\end{equation}
The continuously varying dynamical exponent is therefore
\begin{equation}
\omega
=
-\frac{1}{\xi\ln(1-\rho)}.
\end{equation}

Note that this dynamical exponent depends continuously on both the geometric correlation length, $\xi$ of the percolation cluster and the particle density, $\rho$. As the density increases towards close packing ($\rho\rightarrow1$), the characteristic trapping times become increasingly broad, leading to progressively smaller values of $\omega$ and correspondingly slower relaxation.

Because the underlying lattice is discrete, the full probability density $\psi(t)$ exhibits logarithmically spaced peaks. The $r$-th peak occurs at
\begin{equation}
t_r \approx 
       \left(\frac{1}{1-\rho}\right)^r,
\end{equation}
with its local maximum exponentially suppressed according to the continuous envelope $\psi(t)\sim t^{-\omega}$.

Consequently, the survival probability, defined as the probability that a particle remains trapped in a side branch for at least a time $t$, is obtained by the following summation,
\begin{equation}
    S(t) = \sum_{h=0}^\infty e^{-\lambda_h t} w(h) e^{-h/\xi}
\end{equation}
where $w(h)$ is the weight associated with a particle returning to the backbone from depth $h$.~This yields $S(t) \sim t^{-(1+\omega)}$. To extract the exponent $\omega$ from the simulation data, we consider the scaled survival probability, defined as $tS(t)$. This quantity scales as $t^{-\omega}$, enabling a direct measurement of $\omega$. This heavy-tailed trapping mechanism dictates the macroscopic dynamics of the system. 

{

The return probability, $P_{\mathrm{ret.}}(t)$, is defined as the probability that a tagged particle is found at its initial position after time $t$. This probability receives contributions from two mutually exclusive processes: (i) the particle returns to its original position while remaining within the original side branch $P_{\mathrm{ret.}}^{\mathrm{branch}}(t)$, and (ii) the particle escapes the side branch and diffuses along the backbone, before returning to its initial position $P_{\mathrm{ret.}}^{\mathrm{backbone}}(t)$. We argue below that the long-time behavior is dominated by the first contribution. Combining the distribution of branch depths with the characteristic escape time from depth $h$, we have
\begin{equation}
    P_{\mathrm{ret.}}^{\mathrm{branch}}(t) \sim t^{-\omega}.
\end{equation}
In contrast, particles that escape to the backbone must first return by diffusion along the backbone before returning to their initial position. This contribution is therefore suppressed by a return probability of diffusion along a $d$-dimensional backbone. We therefore have
\begin{equation}
    P_{\mathrm{ret.}}^{\mathrm{backbone}}(t) \sim t^{-(d/2 + \omega)}.
\end{equation}

Consequently, in the long-time limit, the return probability is dominated by the first contribution yielding
\begin{equation}
    P_{\mathrm{ret.}}(t) \sim t^{-\omega}.
\end{equation}
Therefore the return probability is dominated by the not-so-rare events in which the tagged particle remains within a side branch.

\subsection{Threshold Density}

To estimate the threshold density $\rho^*$, we consider the mean waiting time of a particle located at depth $r$ on a side branch. Below $\rho^*$, the mean waiting time is expected to remain finite. We assume that the survival probability of the waiting time at depth $r$ has the asymptotic form
\begin{equation}
S_r(t) \sim e^{-\lambda_r t},
\end{equation}
where the escape rate decreases with depth as
\begin{equation}
\lambda_r \sim (1-\rho)^r.
\end{equation}
Consequently, the mean waiting time for a particle at depth $r$ scales as
\begin{equation}
\langle \tau_r \rangle
= \int_0^\infty S_r(t)~dt
\sim \lambda_r^{-1}
\sim (1-\rho)^{-r}.
\end{equation}

The number of sites at depth $r$ on a side branch decreases exponentially with $r$,
\begin{equation}
N(r) \sim e^{-r/\xi},
\end{equation}
where $\xi$ characterizes the typical branch depth. The overall mean waiting time therefore scales as
\begin{equation}
\langle \tau \rangle
\sim \sum_{r=0}^{\infty}
N(r)\langle\tau_r\rangle
\sim
\sum_{r=0}^{\infty}
e^{-r/\xi}(1-\rho)^{-r}.
\end{equation}
Thus,
\begin{equation}
\langle \tau \rangle
\sim
\sum_{r=0}^{\infty}
\left[
\frac{e^{-1/\xi}}{1-\rho}
\right]^r.
\end{equation}
This geometric series converges only when
\begin{equation}
\frac{e^{-1/\xi}}{1-\rho}<1.
\end{equation}
The threshold density $\rho^*$ is therefore obtained from the convergence boundary,
\begin{equation}
\frac{e^{-1/\xi}}{1-\rho^*}=1,
\end{equation}
which gives
\begin{equation}
\rho^*=1-e^{-1/\xi}.
\end{equation}
Hence, for $\rho<\rho^*$, the mean waiting time remains finite, whereas for $\rho\geq\rho^*$ it diverges within this scaling argument. For example, for a percolation cluster with correlation length $\xi \approx 0.56$, the corresponding threshold density, $\rho^* \approx 0.832$.

%%%%%%%%%%%%%%%%%%%%%%%%%%%%%%%%%%%%%%%%%%%%%%%%%%%
\begin{figure}
    \centering
    \includegraphics[width=1\linewidth]{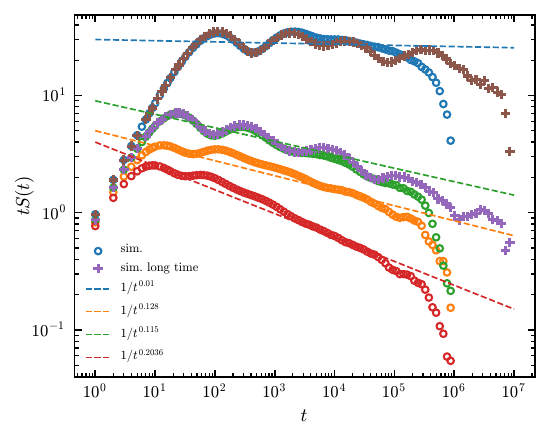}
    \caption{
    Scaled survival probability of particles in the side branches of the irregular comb. Simulations are performed for various particle densities, $\rho$, and exponentially distributed branch lengths, $P(h)\sim e^{-h/\xi}$.
    Data are presented for $T_{\max}=10^6$ (sim.) and $T_{\max}=10^7$ (sim. long time). Note that Toom's dynamics was implemented at the floor of the irregular comb. The dashed lines indicate the expected asymptotic power law behavior, with exponents predicted by $\omega=-1/[\xi \ln(1-\rho)]$.
    }
    \label{fig:exponent_measured}
\end{figure}

%%%%%%%%%%%%%%%%%%%%%%%%%%%%%%%%%%%%%%%%%%%%%%%%%%%

%%%%%%%%%%%%%%%%%%%%%%%%%%%%%%%%%%%%%%%%%%%%%%%%%%%
\begin{figure}
    \centering
    \includegraphics[width=1\linewidth]{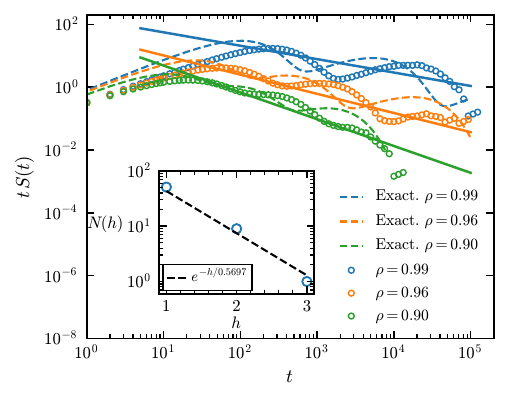}
    \caption{Scaled survival probability distribution measured on a percolation cluster for three different particle densities, exhibiting power-law behavior. The measurements are performed on a spanning cluster of a $30\times 30$ lattice (see the right panel of Fig.~\ref{fig:snap_perc_cluster}). The distribution of side-branch lengths, which are approximately one-dimensional, is fitted to an exponential distribution form, yielding a correlation length $\xi=0.5697$ (see inset; data points show the number of side branches of length $h$, and the solid line denotes the exponential fit). The exponents are determined from $\omega = -1/[\xi \ln(1-\rho)]$. For $\rho=0.90, ~0.96, ~0.99$, we obtain $\omega \approx 0.76,~0.54, ~0.38$, respectively.
    The dotted lines show the exact waiting time distribution calculated from Eq.~\eqref{eq:surv} for the percolation cluster. The log-periodic oscillations in $S(t)$ arise from the discrete nature of the cluster depth. The observed waiting-time distributions are shifted toward larger waiting times relative to the exact calculation, indicating the presence of waiting-time correlations in the system.
    }
    \label{fig:WTD_powerlaw_perc}
\end{figure}

\subsection{Numerical Measurements}
{
 Finally, we test our theoretical predictions by numerically simulating SEP dynamics on three related disordered geometries: a ring with exponentially distributed side branches (irregular comb), the two dimensional torus with exponentially distributed side branches (irregular carpet) and the actual percolation cluster.

In Fig.~\ref{fig:exponent_measured}, we consider a one-dimensional periodic backbone with side branches whose lengths are drawn from an exponential distribution, $\exp(-h/\xi)$. To suppress correlations between successive waiting times, we implement Toom dynamics for particles on the backbone. We find a continuously varying power-law behavior as the particle density and branch-length distribution are varied.  We extend the same construction to an irregular carpet by allowing particles to diffuse along a two-dimensional periodic floor, while retaining the randomly varying branch lengths in Fig.~\ref{fig:peak_comparison} in the appendix. Here, we implement LSEP dynamics along the floor of the carpet to suppress correlations between waiting times. The resulting behavior closely resembles that of an irregular comb, showing that the power-law behavior is primarily associated with the particle density and with the distribution of branch lengths, rather than the dimensionality of the backbone.

In Fig.~\ref{fig:WTD_powerlaw_perc}, we study SEP dynamics directly on a supercritical percolation cluster. In this case, the branch structure and its associated length scales arise naturally from the underlying percolation disorder. The measured power-law behavior is in good agreement with the exponent obtained from our calculation. In this case, we use the standard SEP dynamics without an additional relaxation mechanism for suppressing correlations, since implementing such dynamics on the disordered percolation cluster is technically more involved.
}

The dependence of the dynamical exponent, $\omega$, on both the particle density $\rho$ and the correlation length $\xi$ is detailed in Appendix Fig.~\ref{fig:power_law_summary}. Furthermore, we demonstrate the robustness of this scaling behavior by computing the probability distribution across various maximum depth cutoffs, confirming that the asymptotic power-law tail remains stable against finite-size truncation (Appendix Fig.~\ref{fig:power_law_summary}). The log-periodic oscillations in the survival probability are coming from the discrete nature of the depth parameter $h$. When running the simulation for longer times, we observe many more peaks appearing in the waiting-time distribution, agreeing with the calculated exponent (see Fig.~\ref{fig:exponent_measured}). Note that the time-scaled survival probability curves exhibit the same behavior in one and two dimensions. Furthermore, we systematically examine whether the individual peaks observed in the waiting-time distribution can be predicted quantitatively. To this end, we label particles by the number of particles above them on the side branch, collect the waiting-time statistics for each particle, construct the corresponding cumulative waiting-time distributions, and compare them with the exact calculation (see Appendix, Fig.~\ref{fig:peak_comparison}).

\section{Discussion and Conclusion}
\label{sec:summary}

In this study, we explored the waiting-time distribution of hard-core interacting particles undergoing the symmetric exclusion process (SEP) in irregular geometries, including a supercritical percolation cluster. We established that the system exhibits strong dynamical heterogeneity~\cite{Berthier_DH,Spatial_het,10.1063/1.4795539} driven by a purely topological bias (see Fig.~\ref{fig:disp_field}). The lattice partitions into distinct mobility classes separated by several orders of magnitude: particles on the backbone diffuse rapidly and dominate the macroscopic mean squared displacement ($\langle r^2 \rangle \sim t$) (see Fig.~\ref{fig:combined_results}(c) in the appendix), while particles pushed into side branches experience an exponential hierarchy of trapping times. This geometric crowding produces multiple, logarithmically separated peaks in the waiting-time probability density $\psi(\tau_w)$. Interestingly, we find a continuously varying, density and correlation length dependent-dynamical exponent, $\omega = [-\xi \ln(1-\rho)]^{-1}$. Since the correlation length of the percolation cluster near $p_c$ scales as $\xi = \xi_0 |p-p_c|^{-4/3}$, where the branch-depth decay length is identified with the standard percolation correlation length~\cite{StaufferAharony1992}. The dynamical exponent can then be expressed as,
\begin{equation}
    \omega = -\frac{|p-p_c|^{4/3}}{\xi_0 \ln(1-\rho)},
\end{equation} 
highlighting its explicit dependence on both the particle density $\rho$ and the distance from percolation threshold ($p_c$). Even though the overall transport remains diffusive, microscopic topological trapping gives rise to an anomalous, heavy-tailed survival probability for tagged particles, $S(t) \sim t^{-(1+\omega)}$, and correspondingly modifies the return-to-origin probability, $P_{\mathrm{ret.}}(t) \sim t^{- \omega}$. This slow algebraic relaxation is reflected in the velocity-velocity autocorrelation function of a tagged particle, which develops a pronounced long-time tail and thereby connects the present dynamics to the well known Alder--Wainwright long-time-tail phenomenon~\cite{PhysRevLett.18.988}. In particular, at high densities, the memory of the initial dynamical state of a tagged particle persists over increasingly long times, causing the velocity-velocity auto-correlations to decay algebraically rather than exponentially. In the high-density limit, $\rho \to 1$, particle motion becomes vacancy-mediated while the underlying percolation geometry remains unchanged. While the collective diffusion coefficient of the symmetric exclusion process is well understood~\cite{PhysRevE.90.052108, Mason2023}, the self-diffusion coefficient of a tagged particle in two-dimensions is a subtler quantity. The high-density limit nevertheless admits a natural description in terms of vacancy-mediated motion. Our analysis, at its current level, requires realizations of percolation clusters away from the percolation threshold. At the critical threshold, the distribution of waiting times exhibits qualitatively different behavior: rather than following a power law, it is characterized by a logarithmic distribution of waiting times.

The SEP on a percolation cluster provides a simple model to understand the broad residence time distributions seen in more complicated realistic settings such as porous materials or complex intracellular environments.

There are several directions for future study. A natural challenge in modeling this steady state is the presence of microscopic dynamic correlations; a particle exiting a branch leaves a local density depletion, significantly elevating its probability of immediate re-entry. By measuring the spatial re-entry probability $P(r)$ (Fig.~\ref{fig:correlation_func}) and implementing Lifted SEP (LSEP) dynamics on the backbone, we successfully suppressed these local memory effects, yielding excellent quantitative agreement with our exact single-branch matrix approximation. An exact treatment of these correlations is a challenging problem. Furthermore, extending this microscopic approach to study the trapping statistics of active, self-propelled tracers in disordered landscapes presents a challenging direction for future work, with potential applications to realistic non-equilibrium dynamics of active matter in biological systems.

%%%%%%%%%%%%%%%%%%%%%%%%%%%%%%%%%%%%%%%%%%
%\clearpage

\section{Acknowledgments}
We acknowledge useful discussions with Mustansir Barma, and Rahul Dandekar for a critical reading of the manuscript. A.C. thanks Sushant Saryal for valuable comments. This project was funded by intramural funds at TIFR Hyderabad from the Department of Atomic Energy (DAE), Government of India, under the Project Identification No. RTI 4007. DD's work was supported by the Indian National Science Academy, New Delhi under the grant No. SP/DP/2023/658.

%%%%%%%%%%%%%%%%%%%%%%%%%%%%%%%%%%%%%%%%%%
\appendix
\section{Symmetric exclusion process on ring and carpet}
\label{app: ring_carpet_com}

%%%%%%%%%%%%%%%%%%%%%%%%%%%%%%%%%%%%%%%%%%%%%%%%%%%%%%%%%%%%%%%%
\begin{figure*}
    \centering
    \includegraphics[width=1\linewidth]{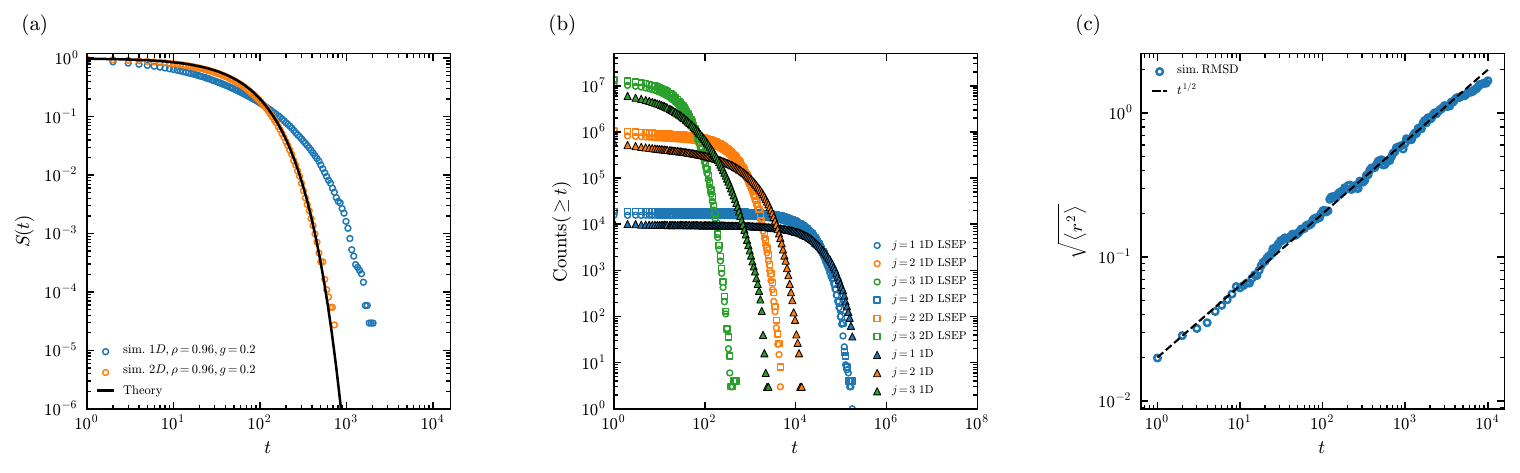}
    \caption{(a) Comparison of the waiting time distribution measured in the carpet and the ring at high particle density ($\rho=0.96$) under a weak external bias ($g=0.2$). Symbols represent simulation results, while the solid line denotes the exact single branch calculation.
    (b) Comparison of the waiting time distributions measured with the standard SEP dynamics and the Lifted SEP dynamics for the ring ($1D$) and carpet ($2D$) geometries.
    (c) Root mean square displacement of tagged particles measured on a supercritical percolation cluster on a lattice of size $60\times 60$ at particle density $\rho=0.99$.
    }
    \label{fig:combined_results}
\end{figure*}

\begin{figure}
    \centering
    \includegraphics[width=0.8\linewidth]{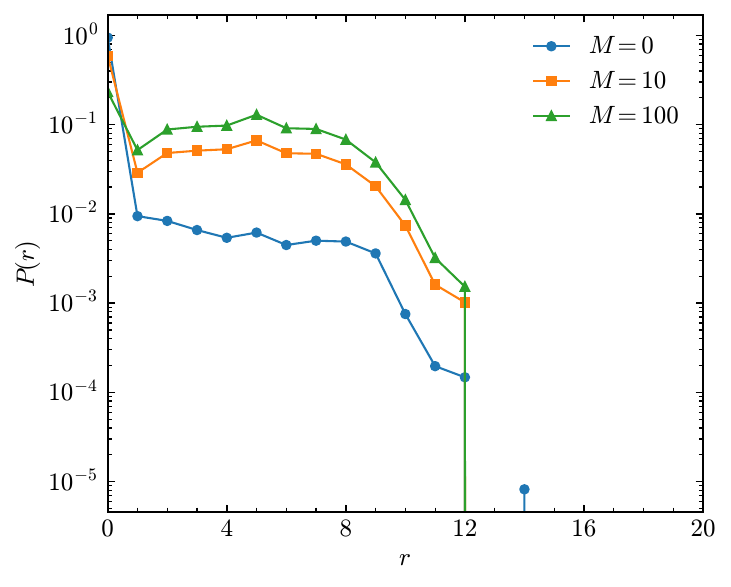}
    \caption{Measured re-entry probability in the carpet for different numbers of Lifted SEP steps, $M$, performed per Monte Carlo update. Increasing $M$ suppresses the re-entry probability at $r=0$, resulting in faster relaxation of correlations.}
    \label{fig:correlation_func}
\end{figure}
%%%%%%%%%%%%%%%%%%%%%%%%%%%%%%%%%%%%%%%%%%%%%%%%%%%%%%%%%%%%%%%%

The single branch method predicts the waiting time distribution independently of the dimensionality of the floor. In this appendix, we compare the single branch calculation with simulation results for the SEP on both the ring and the carpet. Fig.~\ref{fig:combined_results}(a) compares the waiting-time distributions on
the one-dimensional ring and the two-dimensional carpet in the presence
of a weak external bias, $g=0.2$. This comparison illustrates that the
qualitative trapping behavior is similar in the two geometries. Figs.~\ref{fig:combined_results}(b) and (c) focus on the unbiased SEP and on the effect of the
relaxation protocols used to suppress correlations between successive
branch visits.
The agreement between the single branch calculation and the measured distribution is significantly better for the carpet than for the ring.

At higher particle densities, noticeable deviations emerge between the exact calculation and the simulation results, with the discrepancy being more pronounced in the ring geometry. This difference arises because particle motion is more strongly correlated on the ring than on the carpet~\cite{3312643e-a70a-397f-b33f-8297a32239ae,PhysRevA.8.3050,PhysRevB.18.2011}, where the higher-dimensional floor facilitates relaxation of these correlations. Upon implementing the relaxation protocols, the agreement between the measured and calculated waiting time distributions improves substantially for both geometries (Fig.~\ref{fig:combined_results}(b)).

%%%%%%%%%%%%%%%%%%%%%%%%%%%%%%%%%%%%%%%%%%%%%%%%%%%%%%%%%%%%%%%%
\begin{figure*}
    \centering
    \includegraphics[width=1\linewidth]{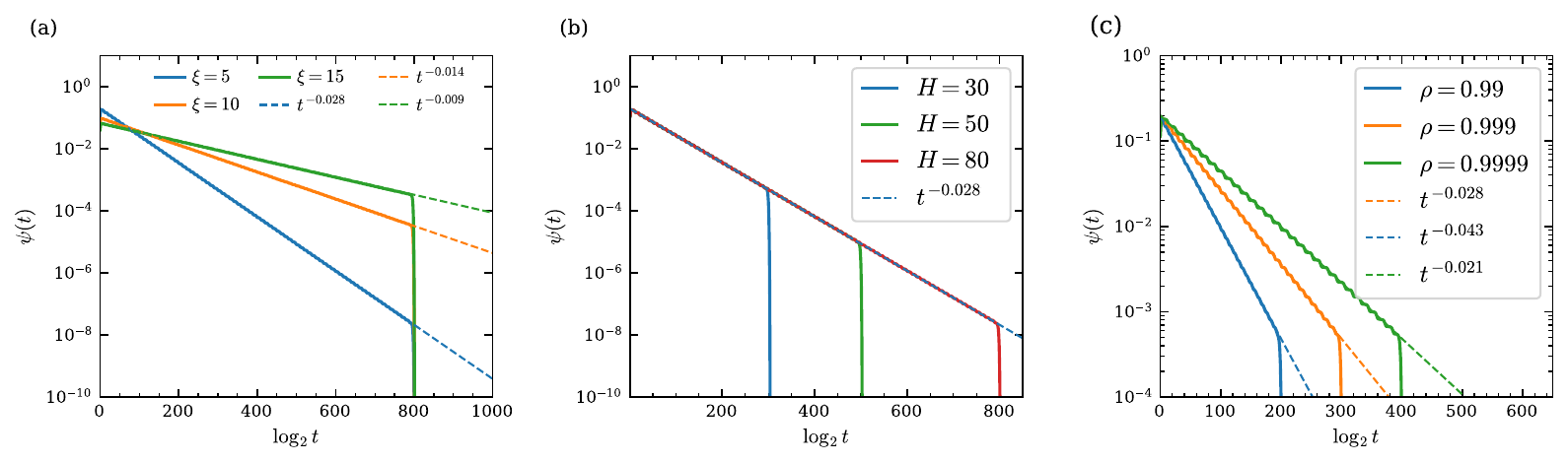}
    \caption{Probability density of the waiting time obtained from the branch model.
    (a) Waiting time distribution for a fixed maximum branch depth, $H_{\max}=80$, and particle density $\rho=0.999$, for three different correlation lengths, $\xi=5, 10, \text{ and } 15$. 
    (b) Waiting time distribution for a fixed correlation length, $\xi=5$, and particle density $\rho=0.999$, for three maximum branch depths, $H_{\max}=30, 50, \text{ and } 80.$
    (c) Waiting time distribution for a fixed maximum branch depth, $H_{\max}=30,$ and correlation length, $\xi=5$, for particle densities $\rho=0.99, 0.999, \text{ and }0.9999$. The dashed line indicates the theoretical power law exponents predicted by $\omega = -\frac{1}{\xi \ln(1-\rho)}$.
    }
    \label{fig:power_law_summary}
\end{figure*}
\begin{figure}[t!]
    \centering
    \includegraphics[width=1\linewidth]{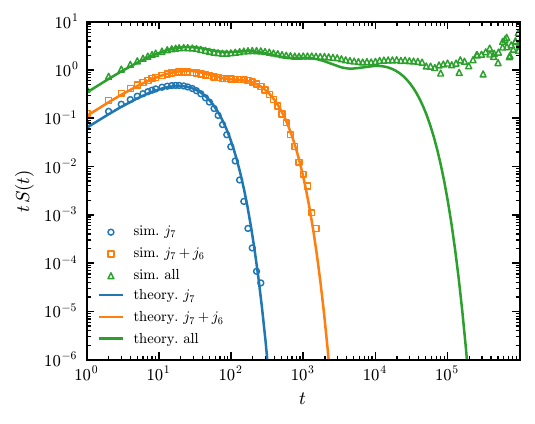}
    \caption{Scaled survival probability distributions conditioned on the occupancy of the side branch. Symbols denote the simulation results from irregular carpet, while the solid line represents the corresponding single branch calculations. Here, $j_n$ denotes the contribution from configurations in which the side branch contains at least $n$ particles. The label $j_7$ corresponds to the survival probability conditioned on the branch containing at least seven particles, while $j_7 + j_6$ denotes the cumulative contribution obtained by summing the conditional survival probabilities for branches containing at least six and at least seven particles. Similarly, `all' denotes the sum over all conditional contributions. Note that LSEP dynamics was implemented on the carpet floor.
    }
    \label{fig:peak_comparison}
\end{figure}
%%%%%%%%%%%%%%%%%%%%%%%%%%%%%%%%%%%%%%%%%%%%%%%%%%%%%%%%%%%%%%%%

We also measure the re-entry probability of a tagged particle on the carpet, both with and without relaxation protocols (Fig.~\ref{fig:correlation_func}). The re-entry probability is defined as the probability that a particle, after leaving a branch and moving along the floor, re-enters a branch after traversing a distance $r$. This quantity provides a simple measure of the spatial correlations in the particle dynamics and enables us to assess the effectiveness of the relaxation protocols in suppressing these correlations.
%%%%%%%%%%%%%%%%%%%%%%%%%%%%%%%%%%%%%%%%%%%%%%%%%%%%%
\begin{figure}[t!]
    \centering
    \includegraphics[width=1\linewidth]{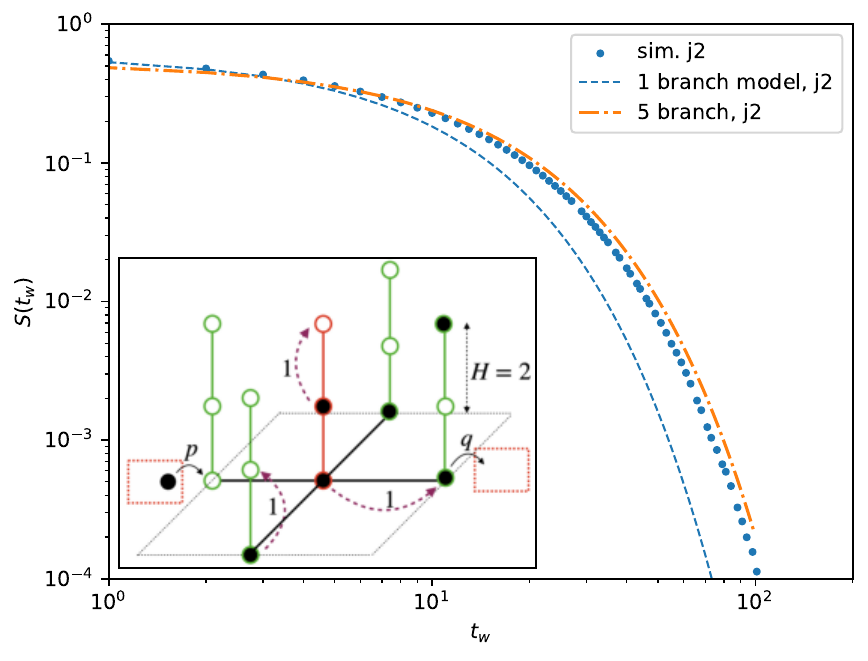}
    \caption{Conditional survival probability measured on the carpet (with side branch of depth $H=2$) with zero bias, compared with the single-branch calculation (1 branch model (Eq.~\eqref{eq:surv_prob}) and five-branch model calculation (5 branch) obtained by implementing Eq.~\eqref{eq:surv_prob} for the five-branch model. The label $j2$ represents the conditional survival probability that at least $2$ particles survive in the side branch up to time $t$. \textbf{Inset}: Schematic representation of the five-branch model with branch depth $H=2$. The central (red) branch is coupled to four identical side branches, each connected to an independent particle reservoir (red dotted box) of density $\rho$. Particles enter and leave the system from the reservoirs with rates $p=\rho$ and $q=1-\rho$, respectively. Circles represent lattice sites, with at most one particle allowed per site. Black circles denote occupied sites. Dotted arrows indicate particle hopping between neighboring sites at unit rate, subject to the hard-core exclusion constraint. For clarity, only a subset of the allowed hopping processes is shown.}
    \label{fig:five_branch_model}
\end{figure}
%%%%%%%%%%%%%%%%%%%%%%%%%%%%%%%%%%%%%%%%%%%%%%%%%%%%%

%%%%%%%%%%%%%%%%%%%%%%%%%%%%%%%%%%%%%%%%%%%%%%%%%%%%

\section{Multi-branch model}
The single branch approximation accurately describes the waiting time distribution over a broad range of densities. However, at very high particle densities, the measured waiting time distribution deviates significantly from the exact calculation. As discussed in the previous section, this discrepancy originates from correlations in the particle dynamics that are neglected in the single branch description. By introducing modified dynamics incorporating long-range jumps, while preserving the equilibrium measure of the SEP, we demonstrated that suppressing these correlations substantially improves the agreement between calculation and simulation.

This naturally raises the question of whether the same improvement can be achieved without modifying the microscopic dynamics. In this section, we introduce a multi-branch model that explicitly incorporates correlations arising from repeated re-entry into neighboring branches. Instead of treating an isolated branch connected to a particle reservoir, the model considers several neighboring side branches coupled through a common reservoir (Fig.~\ref{fig:five_branch_model}), thereby enlarging the state space to account for re-entry-induced correlations. To calculate the survival probability in the five-branch model, we sum over all the possible states of the four side branches while keeping the central branch explicit. The central branch is distinguished only because the survival probability is evaluated on it; in the full carpet, any branch can be regarded as the central branch of an equivalent five-branch cluster formed by its four nearest-neighbor branches.

We compare the survival probabilities calculated in the single-branch and multi-branch model with simulation results obtained for the carpet geometry (Fig.~\ref{fig:five_branch_model}). Although the enlarged configuration space makes the diagonalization of the transition matrix $\mathbf{W}$ computationally more demanding, the multi-branch model provides a substantially improved prediction of the waiting time distribution without requiring any relaxation protocols.

To clarify how the single-branch framework accounts for side-branch length fluctuations, we consider an irregular carpet where the branch depths $h$ attached to the two-dimensional floor are drawn from an exponential distribution $P(h) \sim e^{-h/\xi}$. Figure~\ref{fig:peak_comparison} shows the scaled survival probability $t S(t)$ resolved by the occupancy state of the branch. The label $j_n$ denotes the conditional survival probability for configurations containing at least $n$ particles trapped within the branch. 
By taking cumulative sums over these conditional contributions ($j_7$, $j_7 + j_6$, up to `all'), the individual discrete steps smooth out into the overarching power-law envelope $t S(t) \sim t^{-\omega}$. The agreement between the exact single-branch calculations (solid lines) and the irregular carpet simulation data (symbols) demonstrates that the global power-law dynamics on a spatially disordered lattice arise directly from the superposition of these conditionally trapped microstates.

%\clearpage

\bibliographystyle{apsrev4-2}
\bibliography{2d_SEP_Bibliography}

\end{document}